\documentclass[11pt]{article}
\usepackage{verbatim}       
\usepackage{amsfonts}       
\usepackage{amsmath}        
\usepackage{bm}         
\usepackage{epsfig}

\def\ben{\begin{equation}}
\def\een{\end{equation}}

\let\a=\alpha   \let\d=\delta

\let\w=\omega

\let\pa=\partial
\def\ba{\begin{array}}
\def\ea{\end{array}}

\def\dalemb#1#2{{\vbox{\hrule height .#2pt
        \hbox{\vrule width.#2pt height#1pt \kern#1pt
                \vrule width.#2pt}
        \hrule height.#2pt}}}

\newcommand{\bea}{\begin{eqnarray}}
\newcommand{\eea}{\end{eqnarray}}

\def\be{\begin{equation}}
\def\ee{\end{equation}}
\def\bea{\begin{eqnarray}}
\def\eea{\end{eqnarray}}

\def\a{\alpha}
\def\d{\delta}

\def\w{\omega}
\def\ep{{\epsilon}}

\def\pa{\partial}

\def\cale{{\mathcal E}}
\def\calb{{\mathcal B}}
\def\calp{{\mathcal P}}

\begin{document}
\begin{flushright}
NSF-KITP-07-145 \\
arXiv:0706.3228 [hep-th]
\end{flushright}

\begin{center}
\vspace{1cm} { \LARGE {\bf Ohm's Law at strong coupling: \\ S duality and the cyclotron resonance}}

\vspace{1.1cm}

Sean A. Hartnoll$^\flat$ and Christopher P. Herzog$^\sharp$

\vspace{0.8cm}

{\it $^\flat$ KITP, University of California\\
     Santa Barbara, CA 93106-4030, USA }

\vspace{0.8cm}

{\it $^\sharp$ Physics Department, University of Washington \\
     Seattle, WA 98195-1560, USA }

\vspace{0.8cm}

{\tt hartnoll@kitp.ucsb.edu, herzog@u.washington.edu} \\

\vspace{2cm}

\end{center}

\begin{abstract}
\noindent
We calculate the electrical and thermal
conductivities and the thermoelectric coefficient of a class of
strongly interacting 2+1 dimensional conformal field theories with
anti-de Sitter space duals. We obtain these transport coefficients
as a function of charge density, background magnetic field,
temperature and frequency. We show that the thermal conductivity
and thermoelectric coefficient are determined by the electrical
conductivity alone. At small frequency, in the hydrodynamic limit,
we are able to provide a number of analytic formulae for the
electrical conductivity. A dominant feature of the conductivity is
the presence of a cyclotron pole. We show how bulk electromagnetic
duality acts on the transport coefficients.

\end{abstract}

\pagebreak
\setcounter{page}{1}

\section{Introduction}

As was recently pointed out by \cite{HKSS}, the AdS/CFT correspondence
 \cite{jthroat, GKP, EW} may be useful for studying transport properties
 of real world 2+1 dimensional systems at their quantum critical points.
 Phase transitions between quantum Hall states, superfluid-insulator
 transitions in thin films, and magnetic ordering transitions of Mott insulators and
 superconductors are all believed to be examples of
 quantum phase transitions \cite{Sachdevbook}, in which fluctuations
 are driven by the quantum mechanical zero point energy of the system
 rather than the temperature.  Moreover, it is often the case that
 the effective field theory description of the quantum critical point is
 strongly interacting.  Conveniently, the AdS/CFT is a duality that
 provides access to certain
 strongly interacting conformal
 field theories (CFTs) through a classical dual gravitational description in an
 asymptotically anti-de Sitter (AdS) spacetime.\footnote{%
  These CFTs typically have a parameter $N$ which counts the
  number of degrees of freedom and needs to be kept parametrically large.
 }

 In this paper, we use the AdS/CFT correspondence to study the thermal
 and electrical transport properties of a set of strongly interacting
 2+1 dimensional conformal field theories in a background magnetic field $B$
 and at finite charge density $\rho$.
 One example to which our discussion applies is the infrared conformal fixed
 point of maximally supersymmetric $SU(N)$ Yang-Mills theory at large $N$.
 This CFT has 16 supersymmetries and a global $SO(8)$ R-symmetry group.
 The CFT is also believed to describe the low energy dynamics of a set of
 $N$ M2-branes, and hence we will often refer to it as the M2-brane theory.
 The magnetic field we turn on belongs to a $U(1)$ subgroup of the $SO(8)$.
 In general, our results will apply to any CFT with an AdS/CFT dual that may be
 truncated to Einstein-Maxwell theory with a negative cosmological constant:
 \be
 \label{EMaction}
 S = -\frac{1}{2\kappa_4^2} \int d^4x \sqrt{-g} \left[ R - L^2 F_{\mu\nu} F^{\mu\nu} + \frac{6}{L^2} \right]
 \ee
with $L$ the radius of curvature of $AdS_4$ and $\kappa_4$ the gravitational coupling.

It was observed in \cite{Hartnoll:2007ai} that by placing an
electrically and magnetically charged black hole in the center of
$AdS_4$, we can study the dual CFT at finite temperature $T$,
charge density $\rho$, and magnetic field $B$.\footnote{%
 The thermodynamic properties of this M2-brane theory at
 nonzero $\rho$ and $T$ but zero $B$ were investigated
 in \cite{Cai:1998ji,Chamblin:1999tk,CveticGubser}.
}
The AdS/CFT
dictionary maps fluctuations in the gauge potential $A_{\mu}$ and
metric $g_{\mu\nu}$ to the behavior of a conserved current $J^\mu$
and the stress tensor $T^{\mu\nu}$ in the 2+1 dimensional CFT. In
particular, the dictionary provides a way of calculating two-point
correlation functions of $J^\mu$ and $T^{\mu\nu}$. From these
two-point functions, linear response theory allows one to extract
the thermal and electrical transport coefficients of the CFT. The
formalism for these finite temperature AdS/CFT calculations was
first worked out in \cite{SS, HS}.

In addition to \cite{HKSS}, there have been a handful of
earlier studies of the transport properties of this M2-brane theory.
Ref.~\cite{Herzoghydro} calculated the viscosity and R-charge diffusion constants
in the limit $B=\rho=0$, while \cite{Herzog:2003ke} studied sound waves and
\cite{Saremi:2006ep} calculated the viscosity at finite charge density.

In this paper, extending work of Ref.~\cite{Hartnoll:2007ai}, we
study the electrical conductivity  {\boldmath\mbox{$\sigma$}}, the
thermoelectric coefficient {\boldmath\mbox{$\alpha$}} and the
thermal conductivity  {\boldmath\mbox{$\bar \kappa$}}. In the
presence of a magnetic field, these three quantities are in
general $2\times 2$ antisymmetric matrices $M$ with $M_{xx} =
M_{yy}$ and $M_{xy} = -M_{yx}$. The constraints are due to
rotational
 invariance.  At the level of linear response, we have
 \be\label{eq:transport}
 \left(
 \begin{array}{c}
 \vec J \\
 \vec Q
 \end{array}
 \right)
 =
 \left(
 \begin{array}{cc}
 {\boldmath\mbox{$\sigma$}} & {\boldmath\mbox{$\alpha$}} \\
{\boldmath\mbox{$\alpha$}} T &{\boldmath\mbox{$\bar \kappa$}}
 \end{array}
 \right)
 \left(
 \begin{array}{c}
 \vec E \\
 -\vec \nabla T
 \end{array}
 \right) \ .
 \ee
 Here $\vec \nabla T$ is the temperature gradient, $\vec E$ the applied electric field,
 $\vec J$ the electrical current, and $\vec Q$ the heat current.  We allow for
 $\vec \nabla T$ and $\vec E$ to have a time dependence of the form $e^{-i \omega t}$.

Defining
\be
\sigma_\pm = \sigma_{xy} \pm i \sigma_{xx}  \; ; \; \; \;
\hat \alpha_\pm = \alpha_{xy} \pm i \alpha_{xx} \; ; \; \; \;
\bar \kappa_\pm  = \bar \kappa_{xy} \pm i \bar \kappa_{xx} \ ;
\ee
in Section \ref{sec:rels} we demonstrate the following relations using the AdS/CFT
dictionary:
\begin{eqnarray}
\label{WFrelone}
\pm \hat \alpha_\pm T \omega &=&  ( B \mp \mu \omega) \sigma_\pm - \rho\ , \\
\pm \bar \kappa_\pm T \omega
\label{WFreltwo}
&=& \left(\frac{B}{\omega} \mp \mu \right) \hat \alpha_\pm T
\omega - sT + m B
\ .
\end{eqnarray}
We have introduced the chemical potential $\mu$, the magnetization $m$, and
the entropy density $s$. Thus the problem
is reduced to computing the electrical conductivity $\sigma_\pm$,
which is then the focus of this paper.

We think it likely that
(\ref{WFrelone}) and (\ref{WFreltwo}) hold very generally.  They can
be derived from Ward identities \cite{Yaffenotes} which will
hold true for any theory with a hydrodynamic limit
in which gravitational and electromagnetic self-interactions can be ignored.
A possible source of confusion in interpreting our results is the nondynamical
nature of the 2+1 dimensional electromagnetic fields.  We work
in a limit where the magnetic and electric fields in the sample
are imposed externally and the plasma itself does not contribute to the
electromagnetic field.  This limit is imposed on us by the AdS/CFT formalism
where we calculate correlation functions of global currents which can only
be thought of as very weakly gauged.

We are able to calculate $\sigma_\pm$ in a number of different limits, extending previous
results \cite{HKSS, Hartnoll:2007ai}.
In \cite{HKSS}, the electrical conductivity at $B=\rho=0$ was observed to be a constant
independent of the frequency of the applied electric field:
\be
\sigma_{xx} = \frac{1}{g^2} \equiv \frac{2L^2}{\kappa_4^2} \ .
\ee
In \cite{Hartnoll:2007ai},
the d.c. conductivity of the CFT at $B \neq 0$ and $\rho \neq 0$ was shown
to give rise to the Hall effect
\be
\sigma_{xy} = \frac{\rho}{B} \ .
\ee

We consider two limits.  In Section \ref{sec:smallBrho}, we consider
 small $\omega$, $B$ and $\rho$ with
$B^2 / \omega s^{3/2}$ and $\rho^2/ \omega s^{3/2}$ held fixed while in
Section
\ref{sec:smallB} we keep only $\omega$ and $B$ small with
$B/\omega s^{1/2}$ fixed. From these limits, we can reconstruct
the complexified conductivity
\be
\label{MHDsigma}
\sigma_+ = i \sigma_Q \frac{\omega + i \omega_c^2/\gamma + \omega_c}
{\omega + i \gamma - \omega_c} \ ,
\ee
where
\be
\label{cyclotronpole}
\omega_c = \frac{B \rho}{\epsilon+\calp} \; , \; \; \; \gamma = \frac{\sigma_Q B^2}{\epsilon+\calp} \ ,
\ee
$\calp$ is the pressure, $\epsilon$ the energy density and
\be\label{eq:sigmaq}
\sigma_Q = \frac{(sT)^2}{(\epsilon+\calp)^2} \frac{1}{g^2} \ .
\ee
The pole at $\omega = \omega_c - i \gamma$ corresponds to a
damped, relativistic cyclotron mode. In components, the
conductivity is
\begin{eqnarray*}
\sigma_{xx} &=&
\sigma_Q \frac{ \omega (\omega + i \gamma + i \omega_c^2/\gamma)}{(\omega+i \gamma)^2 - \omega_c^2} \ , \\
\sigma_{xy} &=& -\frac{\rho}{B} \frac{-2 i \gamma \omega + \gamma^2 + \omega_c^2}{(\omega+ i \gamma)^2 - \omega_c^2} \ .
\end{eqnarray*}
We should emphasize that because in both limits $B$ is held small, these formulae
will in general have subleading corrections in $B$.

In addition to studying $\sigma_\pm$ in various limits analytically,
in Section \ref{sec:numerics} we present numerical results for
arbitrary $B$, $\rho$, and $\omega$. These numerical results match
our analytic results in the appropriate small $B$, $\rho$, and
$\omega$ limits. Furthermore, we exhibit an interesting pattern of
zeroes and poles in the complex frequency plane.

This paper complements \cite{HKMS} which
derives many of the same results using relativistic magnetohydrodynamics (MHD).
While \cite{HKMS} is intended for a condensed matter audience, this paper
is targeted to the high energy community.

The structure of this paper is as follows. In section
\ref{sec:dyonic} we review the dyonic $AdS_4$ black hole
and recast the equations for perturbations about this background
in terms of convenient complexified variables. In section
\ref{sec:complex} we give a simple way of computing the electrical
conductivity $\sigma_\pm$ from the bulk perturbations. We show that
the bulk $SL(2,{\mathbb{Z}})$ electromagnetic duality acts
naturally on the complexified $\sigma_\pm$. S duality is particularly
interesting here, as it relates the conductivities of the theory
when the values of the background magnetic field and charge
density are exchanged. Section
\ref{sec:rels} then relates the other thermoelectric transport
coefficients to $\sigma_\pm$, as we have just described. The remainder
of the paper computes the electrical conductivity. We obtain
analytic results in the hydrodynamic limit that exactly reproduce
the full nontrivial expectations from relativistic
magnetohydrodynamics \cite{HKMS}. We also give numerical results
for the conductivity at arbitrary frequency, magnetic field and
charge density. In a concluding section we discuss applications of
our results to experiments measuring the Nernst effect in
superconductors, and also open questions.

\section{The dyonic black hole}
\label{sec:dyonic}

\subsection{Fluctuations about the black hole}

The bulk spacetime dual to the 2+1 dimensional CFT with both
charge density and a background magnetic field is a dyonic black
hole in $AdS_4$. This black hole has metric
\be
\frac{1}{L^2} ds^2 = \frac{\a^2}{z^2} \left[-f(z) dt^2 + dx^2 + dy^2\right] +
\frac{1}{z^2} \frac{dz^2}{f(z)} \,,
\ee
and carries both electric and magnetic charge
\be\label{eq:Ffield}
F_0 = h \a^2 dx \wedge dy + q \a dz \wedge dt \,,
\ee
where $q, h$ and $\alpha$ are constants. The function
\be
f(z) = 1 + (h^2 + q^2) z^4 - (1 + h^2 + q^2) z^3 \,.
\ee

The Einstein equations for homogeneous fluctuations (no $x,y$
dependence) about this background were written in
\cite{Hartnoll:2007ai} in terms of the gauge potential $A_a$
and $G_a = \delta g_{t a} \a^{-1} z^2$. The Maxwell equations
follow from the Einstein equations. By enforcing that the
fluctuations have no $x,y$ dependence, the equations governing the
fluctuations $A_t$, $\delta g_{tt}$ and $\delta g_{ab}$ must
decouple from the equations governing $A_a$ and $\delta g_{t a}$
by a parity argument, $x\to -x$ and $y \to -y$. (The fluctuations
with a $z$ index can all be set to zero consistently by a gauge
choice.) The fluctuations $A_a$ and $\delta g_{t a}$ are parity
odd while the remaining fluctuations are parity even; the
equations of motion we consider are linear.  If the fluctuations
have an $x,y$ dependence of the form $e^{ik \cdot x}$, then the
parity odd wave vector $k$ can mix the two fluctuations, but by
assumption we have no such $x,y$ dependence.
We have checked this decoupling explicitly.

The equations are greatly simplified by the following two steps.
First, introduce the electric and magnetic field strengths of the
perturbations
\bea
E_a & = & -(\dot A_a + \alpha h \ep_{ab} G_b )\,, \\
B_a & = & -\alpha f(z) \ep_{ab} A_b' \,.
\eea
Prime denotes differentiation with respect to $z$ and dot, with
respect to $t$. The index $a=x,y$ and $\ep_{x y} = 1$. Second,
introduce the following complex combinations of the $x$ and $y$
components
\bea\label{eq:curly}
\cale_\pm = E_x \pm i E_y \; ; \; \; \;
\calb_\pm = B_x \pm i B_y \ .
\eea
Note that $\cale_-$ and $\calb_-$ are not generally the complex
conjugates of $\cale_+$ and $\calb_+$, as $E_a$ and $B_a$ are
generically complex.

The fluctuations are then described by the following pair of
equations
\bea
\label{EBeqone}
f (q \cale_+ + h \calb_+)' + w (h \cale_+
- q \calb_+) & = & 0 \,, \\
\label{EBeqtwo}
\frac{w}{4 z^2} \left(\cale_+' - \frac{w}{f} \calb_+ \right) + h^2
\calb_+ + q h \cale_+ & = & 0 \,.
\eea
The time dependence has been taken to be $e^{-i \omega t}$ with $w
= \omega/\alpha$. The variables $\cale_-$ and $\calb_-$
obey identical equations, but with $h
\to -h$.

The equations are easily seen to be invariant under
electromagnetic (or S) duality, that is, under: ${\mathcal{E}} \to
{\mathcal{B}}$, ${\mathcal{B}}
\to - {\mathcal{E}}$, $h \to -q$ and $q \to h$. It is also
straightforward to obtain decoupled second order equations for
${\mathcal{E}}$ and ${\mathcal{B}}$; we shall describe these
below.

Let us make this notion of S duality more precise.  In our
asymptotically $AdS_4$  background, electromagnetic duality is
often defined as the action $(2\pi/g^2) F \to {\star} F$ where
$\star$ is Hodge duality and depends on the metric:
\be
\star F \equiv
\frac{\sqrt{-g}}{4} {\epsilon_{\mu\nu\rho\sigma}} F^{\rho \sigma} \, dx^\mu \wedge
dx^\nu \,.
\ee
(The indices of $\epsilon_{\mu\nu\rho\sigma}$ are raised lowered
using the metric and $\epsilon_{1234} \equiv 1$.)  This duality
transformation acts only on $F$ and not on the metric. In our
case, $F = F_0 +
\delta F$.  We have chosen $B_a$ to correspond to the spatial
components of $\delta F$ while $E_a$ corresponds to the spatial
components of $\delta{\star} F$, in both cases multiplied by an
overall factor of $\alpha f(z)$ for convenience.  Because the
Hodge star is metric dependent, part of $E_a$ comes from a metric
fluctuation. In contrast, $B_a$ is metric fluctuation independent.
An important
point is that the electromagnetic duality transformation cannot
change the parity of the fluctuations.  The reason is that
$\epsilon_{\mu\nu\rho\sigma}$ is parity even.

\subsection{Some thermodynamics}

The correspondence between the thermodynamics of the black hole
and the dual field theory was described in \cite{Hartnoll:2007ai}.
The quantities of interest to us here are the following: The
temperature of the field theory is given by
\be\label{eq:temperature}
T = \frac{\a (3 - h^2 - q^2)}{4 \pi} \,.
\ee
The background magnetic field, magnetization, charge density, and
chemical potential are
\be\label{eq:Bandrho}
B = h \a^2 \,, \quad m = - \frac{h \alpha}{g^2} \, , \quad
\rho = - \frac{q \a^2}{g^2} \, , \quad \mbox{and} \quad
\mu = -q \alpha \ .
\ee
Some useful expressions for the entropy density, energy
density and pressure are
\be
s = \frac{ \pi \alpha^2}{g^2} \; , \; \; \;
\epsilon = \frac{\alpha^3}{g^2} \frac{1}{2} (1 + h^2 + q^2) \ ,
\quad \mbox{and} \quad
P = \epsilon/2 + m B \ .
\ee
 The expression given here for $P$ is the derivative
 of the free energy with respect to volume.
 In the Introduction, we used a different pressure,
 \be
 {\mathcal P} = \langle T_{aa} \rangle = \epsilon/2 \ .
 \ee
 These formulae give a (nonlinear) map between the bulk quantities
$\a,h,q$ and the field theory quantities $T$, $B$, $\rho$, $\mu$, $s$,
$m$, $\epsilon$, and $P$.
 For the superconformal fixed point of the maximally supersymmetric
 $SU(N)$ Yang-Mills theory, the bulk coupling $g^2$ can be related to field theory
 variables.  In particular,
 \be
\frac{1}{g^2} = \frac{ \sqrt{2} N^{3/2}}{6 \pi} \,.
\ee

\section{Complex conductivities}
\label{sec:complex}

\subsection{Ohm's Law}

We consider a generalized version of Ohm's Law such as might
govern the linear response of a material in a constant background
magnetic field to a time varying electrical field:
\be
\left(
\begin{array}{c}
J_x \\
J_y
\end{array}
\right)
=
\left(
\begin{array}{cc}
\sigma_{xx} & \sigma_{xy} \\
\sigma_{yx} & \sigma_{yy}
\end{array}
\right)
\left(
\begin{array}{c}
E_x \\
E_y
\end{array}
\right) \ .
\ee
We envision applying a spatially uniform electric field with a
time dependence $\vec E = \vec E_0 e^{-i \omega t}$. The
conductivity tensor {\boldmath\mbox{$\sigma$}} is a $2\times 2$
matrix of complex numbers which are a function of the frequency of
the applied electric field.

We now assume our material possesses rotational invariance. It
follows that $\sigma_{xx} = \sigma_{yy}$ and $\sigma_{xy} = -
\sigma_{yx}$. Ohm's Law becomes
\be
\left(
\begin{array}{c}
J_x \\
J_y
\end{array}
\right)
=
\left(
\begin{array}{cc}
\sigma_{xx} & \sigma_{xy} \\
-\sigma_{xy} & \sigma_{xx}
\end{array}
\right)
\left(
\begin{array}{c}
E_x \\
E_y
\end{array}
\right) \ .
\ee
A more compact representation of Ohm's Law is possible. Let
\be
E_\pm = E_x \pm i E_y  \; ; \; \;
J_\pm = J_x \pm i J_y \ .
\ee
As above, $E_-$ and $J_-$ are not generically the
complex conjugates of $E_+$ and $J_+$. Having introduced $E_\pm$ and $J_\pm$,
we find that Ohm's Law can be written as
\be
J_\pm = \mp i \sigma_\pm E_\pm \,, \qquad  \mbox{where} \qquad
\sigma_\pm = \sigma_{xy} \pm i \sigma_{xx} \ .
\ee

\subsection{Conductivity from the bulk}

We can relate the field theory $E$ and $J$ to the boundary
behavior of a $U(1)$ gauge field in an asymptotically $AdS_4$ bulk
spacetime. There is a direct relation between $E$ and the boundary
value of the bulk electric field ${\mathcal{E}}$ defined in
(\ref{eq:curly})
\be
E_\pm = \lim_{z \to 0} \cale_\pm \,.
\ee
This is the non-normalizable mode of the bulk field giving rise to
a background field in the dual theory. The normalisable bulk mode
gives us a relation between $J$ and ${\mathcal{B}}$
\be
g^2 J_\pm = \lim_{z \to 0} \alpha A_\pm' =  \mp i \lim_{z \to 0} \calb_\pm \,,
\ee
where we used the fact that $f(0) = 1$ and that near the boundary
$A = A^0 + g^2 J z/\alpha + \cdots$. It follows that we can obtain the
conductivity from the bulk as
\be\label{eq:conductivity}
\sigma_\pm = \lim_{z \to 0} \frac{\calb_\pm}{g^2 \cale_\pm}
\,.
\ee

We commented above that the differential equations for
$\cale_+$ and $\calb_+$ are related to those
for $\cale_-$ and $\calb_-$ by sending $h \to -h$. It
follows that
\be
 \sigma_-(h) = - \sigma_+ (-h) \ .
\ee
From this expression we can reconstruct $\sigma_{xx}$ and
$\sigma_{xy}$ from $\sigma_+$ alone, to wit
\begin{eqnarray*}\label{eq:xy}
\sigma_{xy} &=& \frac{1}{2} (\sigma_+(h) - \sigma_+(-h)) \ , \\
\sigma_{xx} &=& \frac{1}{2i} (\sigma_+(h) + \sigma_+(-h)) \ .
\end{eqnarray*}
Similar relations hold for $\hat \alpha$ and $\bar \kappa$.

\subsection{S and T duality}

We noted above that the map $h \to -q$, $q \to h$ may be undone by
letting ${\mathcal{B}} \to - {\mathcal{E}}$ and ${\mathcal{E}} \to
{\mathcal{B}}$. Once we have found the conductivity for given
values of $h$ and $q$, the conductivity for the
electromagnetically (or S) dual values $-q,h$ thus immediately
follows from (\ref{eq:conductivity}).

Furthermore recall that under S duality the bulk electromagnetic
coupling is inverted. More specifically, with an Abelian gauge
theory of the form (\ref{EMaction}), $2\pi / g^2
\to g^2 / 2 \pi$. We implemented this effect above by scaling
the field strength $F$ by $g^2/2\pi$.  Thus, we have
\be\label{eq:duality}
S: \quad 2 \pi \sigma^{(q,h)} = \frac{-1}{2 \pi \sigma^{(h,-q)}} \,,
\ee
where in this subsection we suppress the $+$ index
of $\sigma_+$, $\cale_+$ and $\calb_+$.\footnote{%
Inverting the coupling $g^2 \sim N^{-3/2}$ takes us out of the
supergravity limit where (\ref{EMaction}) is a valid description.
However, this formal S duality invariance has nontrivial
implications for the conductivity as a function of $B$ and $\rho$
that are within the supergravity regime, as we will see below.
Note that $\sigma$ depends on $g$ multiplicatively. While $g$ of
the S dual theory may be too large for supergravity to be valid,
the fact that the linearised equations of motion do not depend on
$g$ explicitly means that we can simply rescale $g$ back to its
original value. Thus the theory at fixed $g$ has a functional
dependence on $\rho$ and $B$ that is constrained by S duality. }

From a field theory point of view, this map implies a rather
nonobvious relation between the theory with background magnetic
field $B$ and charge density $\rho$, and the same theory with
background magnetic field $\rho$ and charge density $B$.
Specifically, the duality acts by
\be\label{eq:S}
S: \quad B \to g^2 \rho  \,, \quad \rho \to - \frac{ g^2 B}{4 \pi^2}
\,, \quad 2\pi \sigma \to \frac{-1}{2 \pi \sigma} \,.
\ee
In obtaining this formula, we have used the fact that the dual
coupling $2\pi/\tilde g^2 = g^2/2\pi$. The AdS/CFT correspondence
implies that this relation must hold for all theories with a
gravity dual described by Einstein-Maxwell theory.

The action $\sigma \to -1/\sigma$ is of course the natural action
of S duality on a complex quantity. One implication of the
formalism we have developed is that the conductivity $\sigma =
\sigma_{xy} + i \sigma_{xx}$ is the correct quantity
to consider insofar as duality is concerned, even when
$\sigma_{xx}$ and $\sigma_{xy}$ themselves are complex.

We can extend the S action on the space of theories to a full
$SL(2,{\mathbb{Z}})$ action as follows
\cite{Witten:2003ya}.\footnote{%
 See also \cite{Leigh:2003ez, deHaro:2007eg, Petkou} for related AdS/CFT
 discussions of this $SL(2, \mathbb{Z})$
 action.
}
 Let us endow the bulk theory with a topological
theta term,
\be
I_{\theta} = \frac{\theta}{8 \pi^2} \int_M F \wedge F \,.
\ee
We have $\theta = 0$ for the dimensional reduction of eleven
dimensional supergravity to Einstein-Maxwell theory. In flux
compactifications to $AdS_4$ there will generally be a nonzero
theta term for the four dimensional gauge fields. The action of T
is simply to let $\theta \to \theta + 2 \pi$. Under this shift,
the action changes by a boundary term
\be
\Delta I = \frac{1}{4\pi} \int_{\pa M} A \wedge F \,,
\label{Tshift}
\ee
where $F = dA$. This shift induces a change in the expectation
value of the dual field theory current through the standard
dictionary
\be
\Delta J_a = \frac{\d \Delta I}{\d A_a(z \to 0)} = \frac{1}{2\pi} \epsilon_{a b} \lim_{z \to 0}
E_b \,.
\ee
In terms of our complexified bulk electric and magnetic field
strengths, this gives (at $z=0$)
\be
\Delta {\mathcal{B}} = \frac{g^2}{2\pi} {\mathcal{E}} \,,
\ee
which, combined with (\ref{eq:conductivity}), immediately gives
the action on the dual conductivity
\be\label{eq:T}
T: \quad 2 \pi \sigma \to 2 \pi \sigma + 1 \,.
\ee
Thus we obtain both the generators S and T of
$SL(2,{\mathbb{Z}})$.
There is also a corresponding shift in the charge density,
$2 \pi \rho \to 2 \pi \rho + B$, under $T$ that comes from
the $A_t F_{xy}$ component of (\ref{Tshift}).

Although it is pleasing to see the electromagnetic
$SL(2,{\mathbb{Z}})$ duality group map cleanly onto the dual
conductivity, we should note that the S and T actions are not on
an equal footing. The S duality gives a relation between the
conductivities of the same theory at specific charge densities and
background magnetic fields. In contrast the T action in the
boundary theory simply involves adding by hand a topological
Chern-Simons term to the CFT action \cite{Witten:2003ya, FradkinKivelson}. Unlike S
duality, it is not a statement about the dynamics of the theory.

\section{Relations between $\sigma$, $\hat \alpha$, and $\bar \kappa$ from AdS/CFT}
\label{sec:rels}

We would like to demonstrate (\ref{WFrelone}) and (\ref{WFreltwo})
using the AdS/CFT dictionary. These relations express the
thermoelectrical conductivity $\hat \alpha$ and the thermal
conductivity $\bar \kappa$ in terms of the electrical conductivity
$\sigma$. For ease of presentation, we will focus on the relations
involving $\sigma_+$ only, and we will often drop the explicit subscript
in this section.

The relation for $\hat
\alpha$ may be derived directly from the bulk equations of motion together
with the definition of the transport coefficients in
(\ref{eq:transport}), so let us do that first. If we define $G =
G_x + i G_y$, then equation (\ref{EBeqtwo}) may be used to obtain
the expectation value of the complexified energy current in terms
of the boundary electric field using the standard holographic
relation between the stress tensor and the boundary expansion
of the metric:
\be
T_t = T_{tx} + i T_{ty}= \lim_{z \to 0} \frac{\alpha G'}{4 z^2
g^2} = \frac{-i (\sigma B - \rho) E}{\w}\,.
\ee
We used the relations $J = -i \sigma E$ and $\displaystyle
\lim_{z \to 0} {\mathcal B} = i g^2 J$ from the previous section, as well as the
definitions of $B$ and $\rho$ in (\ref{eq:Bandrho}).

If we now use the definition of the heat current $Q = T_t - \mu
J$, and the definition of $\hat \alpha$ in the absence of a
thermal gradient, $\vec \nabla T = 0$ in (\ref{eq:transport}), we
obtain
\be\label{eq:alpha}
\hat \alpha T \w = \frac{i T_t \w}{E} - \sigma \mu \w = (B - \mu \omega) \sigma -
\rho\,,
\ee
which is the advertised (\ref{WFrelone}).

A similar argument yields an expression for $\bar \kappa$ in terms
of $\hat \alpha$. However, in this case we need to use a bulk
perturbation that corresponds to a nonzero temperature gradient
$\vec \nabla T$ in field theory. The easiest way to achieve such a gradient is
with a different bulk mode than the one we are considering in the
rest of this paper. Concretely, the following is a pure gauge
solution to the bulk Einstein-Maxwell equations linearised about
the dyonic black hole background
\be\label{eq:puregauge}
\frac{z^2 \delta g_{tt}}{\a} = 2 \omega f(z) \,, \quad \frac{z^2 \delta g_{tx}}{\a} =
- k f(z) \,, \quad A_t = - q \omega (z-1) \,, \quad A_x = q k
(z-1)
\,.
\ee
Here we have dropped an overall space and time dependence $e^{-i
\w t + i k x}$ in all the terms. From this solution one can read
off the boundary currents and electric field
\be\label{eq:puregauge2}
T_t = - \frac{3 k \epsilon}{2 \alpha} \,, \qquad J = -
\frac{\rho k}{\alpha} \,, \qquad
E = - \frac{i k B - i k \omega \mu}{\alpha} \,.
\ee
Note that there is an extra statistical contribution to $E$ relative to our
previous electromotive expressions due to a spatially varying chemical potential
coming from $A_t$ in (\ref{eq:puregauge}): $\delta \mu = - \mu w
e^{-i \w t + i k x}$ leading to a $\Delta E = -\partial_x \delta \mu$.
The $\delta g_{tt}$ term in (\ref{eq:puregauge})
leads to a temperature gradient, but we will not need to evaluate
this explicitly.

We can eliminate $\vec \nabla T$ from (\ref{eq:transport}) to
obtain the following expression for the complexified $\bar \kappa$
\be
\bar \kappa = \hat \alpha \frac{i (T_t - \mu J) - \hat \alpha T E}{i J - \sigma
E}\,.
\ee
Plugging in the expressions (\ref{eq:puregauge2}) and using our
previous result (\ref{eq:alpha}) for $\hat \alpha$ leads to the
result
\be\label{eq:kappa}
\bar \kappa T \omega
= \left(\frac{B}{\omega} - \mu \right) \hat \alpha T
\omega - sT + m B
\ .
\ee
This is our second advertised result (\ref{WFreltwo}). In deriving
this expression, we used the fact that $3 \epsilon/2 = sT + \mu
\rho - m B$.

\subsection{The Ward identity approach}

An alternative approach to these formulae is possible, which
proceeds via Ward identities for the two point functions of the
electric and heat current correlators. These can be derived either
directly from the field theory path integral \cite{Yaffenotes} or using AdS/CFT.
Combining these arguments with the argument above gives the direct
and well established
connection between transport coefficients and retarded Green's
functions at arbitrary frequency.

For the AdS/CFT derivation, we start with the boundary action
derived by \cite{Hartnoll:2007ai}:
\be\label{eq:action}
S_{bry} = \frac{\alpha}{g^2} \int dt d^2 x \left[ - \frac{1}{4}
(1+h^2+q^2) G_a G_a + \frac{q}{2} A_a G_a - \frac{1}{8z^2} G_a
{G_a}' + \frac{1}{2} A_a {A_a}' \right] \ .
\ee
At the boundary, we must be able to express $A'$ and $G'$ as linear combinations of the
boundary values of $A$ and $G$:
\begin{eqnarray*}
A'(0) &=& a A(0) + b G(0)\ ,  \\
G'(0) &=& (3z^2) (c A(0) + d G(0)) \ ,
\end{eqnarray*}
where $A = A_x + i A_y$ and $G = G_x + i G_y$.
The constants $a$
and $b$ are determined from the definition of the conductivity
$\sigma$ in (\ref{eq:conductivity}). We have
\be
\alpha A'(0) = - i \calb(0) = - i \sigma g^2 \cale(0) = \alpha \sigma g^2 (w
A(0) + h G(0)) \,,
\ee
from which it follows that
\be
a = g^2 \sigma w \, ; \quad b = g^2 \sigma h \,.
\ee

To obtain $c$ and $d$, we rewrite (\ref{EBeqtwo}) in terms of $A$
and $G$, yielding
\be
w G'  + 4 z^2 ( h f A' + q ( w A + h G) )  = 0 \ .
\ee
By inserting our expansions for $A'$ and $G'$ on the boundary, we may deduce that
\be
c = - \frac{4}{3} \left( \frac{h}{w} a + q \right) \; ; \; \; \;
d =  \frac{h}{w} c \ .
\ee

Consider the following retarded two-point functions:
\begin{eqnarray*}
G^R_{ab} (\omega) &=& -i \int d^2x dt e^{-i \omega t} \theta(t) \langle [J_a(t), J_b(0)] \rangle \ , \\
G^R_{a \pi_b}(\omega) &=& -i \int d^2x dt e^{-i \omega t} \theta(t) \langle [J_a(t), T_{tb}(0)] \rangle \ , \\
G^R_{\pi_a \pi_b}(\omega) &=& -i \int d^2x dt e^{-i \omega t} \theta(t) \langle [T_{ta}(t), T_{tb}(0)] \rangle \ ,
\end{eqnarray*}
from which we construct the following complexified quantities\footnote{%
One can disentangle the various factors of $i$ and -1 by starting with the
result from linear response theory that
$J_x = G_{xx}^R(\omega) A_x = -G_{xx}^R(\omega) i E_x / \omega$
which means that $\sigma_{xx} = -i G_{xx}^R / \omega$.
}
\be
\langle J J \rangle  = G^R_{xx}- i G^R_{xy} \; , \; \; \;
\langle J T \rangle = G^R_{x \pi_x} - i G^R_{x \pi_y} \; , \; \; \;
\langle T T \rangle = G^R_{\pi_x \pi_x} - i G^R_{\pi_x \pi_y} \ .
\ee
It follows from our expressions for $A'(0)$ and $G'(0)$ and the
boundary action (\ref{eq:action}) that $\langle JJ \rangle =
\omega \sigma_+$ and
\be
\omega \langle J T \rangle_u   = B \langle JJ \rangle_u  - \rho \omega \; ; \; \; \;
\omega \langle T T \rangle_u = - \epsilon \omega + B \langle J T \rangle_u \ .
\label{unsubtractedWard}
\ee
We put a subscript $u$ for unsubtracted on our two-point functions.
The two-point functions generated by the AdS/CFT dictionary may differ
by contact terms from retarded Green's functions.\footnote{%
 These unsubtracted contact terms are a general feature of
 correlation functions derived from generating functionals.
 Already in the case without a chemical potential
 and a magnetic field they are present as can be seen
 from setting $\langle JT \rangle_u$ to zero in (\ref{unsubtractedWard}) and
 was noticed in \cite{Herzog:2003ke, Policastro:2002tn}.
}
By definition, the retarded Green's functions should vanish in the $\omega \to 0$ limit, but it is not a priori clear that our $\langle J J \rangle_u$,  etc.~will.

Fortunately, \cite{Hartnoll:2007ai} calculated these two-point functions in the
$\omega \to 0$ limit for this M2-brane theory
and we can use their results to establish the contact terms.
We have
\be
 \langle JJ \rangle_u = \frac{\rho}{B} \omega + {\mathcal O}(\omega^2)\; ; \; \; \;
 \langle JT \rangle_u = \frac{3 \epsilon}{2 B} \omega + {\mathcal O}(\omega^2)  \ .
\ee
Thus we see that $\langle JJ \rangle_u = \langle JJ \rangle$ and
$\langle JT \rangle_u = \langle JT \rangle$.  However, we find from (\ref{unsubtractedWard}) that
\be
\lim_{\omega \to 0} \langle TT \rangle_u = \frac{\epsilon}{2} \ .
\ee
Thus, we define $\langle TT \rangle \equiv \langle TT \rangle_u - \calp$, yielding the
generalized Ward identities
\be
\omega \langle J T \rangle   = B \langle JJ \rangle  - \rho \omega \; ; \; \; \;
\omega \langle T T \rangle = - \epsilon \omega - \calp \omega+ B \langle J T \rangle \ .
\ee
Note that from the structure of the AdS/CFT generating functional for these correlation functions,
the Onsager type relation $\langle JT \rangle = \langle TJ \rangle$ follows.

We are really
interested in the two-point functions involving not $T_{at}$ but
the heat current $Q_a = T_{at} - \mu J_a$.  We find that
\begin{eqnarray}
\langle J Q \rangle &=& (B- \mu \omega)  \sigma_+ - \rho \ ,
\label{JQsigmaid} \\
\langle Q Q \rangle
&=&  \left( \frac{B}{\omega} - \mu\right)\langle J Q
\rangle-\epsilon-\calp + \mu \rho \ .
\label{JQQQid}
\end{eqnarray}
Naively, the principle of linear response tells us to define
\be
\omega \hat \alpha_+ T \equiv \langle JQ \rangle \; ; \; \; \;
\omega \bar \kappa_+ T \equiv \langle QQ \rangle \ ,
\ee
which recovers precisely (\ref{eq:alpha}) and (\ref{eq:kappa}).

\subsection{Magnetization subtractions}

There is a
subtlety associated with magnetization currents which we have thus
far not addressed. The transport coefficients are often defined
not with respect to the total charge and heat currents which we
have called $J$ and $Q$, but to the transport currents, from which
the divergence free magnetization currents have been subtracted:
\[
\vec J_{tr} = \vec J - \vec \nabla {\times} \vec m \; ; \; \; \;
 ({\vec T_{tr}})_{t } = \vec T_t - \vec \nabla {\times} \vec m^E \ .
\]
Here $m^E$ is energy magnetization density. For d.c. currents in
the dyonic black hole theory, it is shown in \cite{HKMS} that $m^E
= \mu m/2$.
In \cite{HKMS}, these subtractions
are performed, and to compare with the results there, we need
to consider these subtractions here as well.

In the $\omega \to 0$ limit, Refs.~\cite{HKMS,
Cooperetal} showed that these subtractions lead to the following
modification of the relation between $\hat \alpha$ and $\bar
\kappa$ and the retarded Greens function for our theory:
\be
\omega \underline{ \hat \alpha}_+ T \equiv \langle JQ \rangle + m \omega \; ; \; \; \;
\omega \underline {\bar \kappa}_+ T \equiv \langle QQ \rangle  - \mu m \omega \ .
\ee
We use an underscore to denote the transport coefficients after
subtracting the effect of magnetization currents. Using the
thermodynamic identity $\epsilon + P = s T +
\mu
\rho$, (\ref{JQQQid}) becomes a little simpler
\be
\omega \underline{ \bar \kappa}_+ T = \left(\frac{B}{\omega} - \mu \right) \omega
\underline{\hat \alpha}_+ T - s T  \ .
\label{WFrelonemod}
\ee
To our knowledge, the theory of magnetization subtractions at
finite frequency has not been developed.  We simply note that if
we insist on keeping the relation (\ref{eq:alpha}) between
$\underline{\hat \alpha}$ and $\sigma$ the same, and this relation
appears to be consistent with magnetohydrodynamics \cite{HKMS},
then we need to make a corresponding frequency dependent
magnetization subtraction from $\sigma$:
\be
\underline {\sigma}_+ = \sigma_+ + \frac{m \omega}{B - \mu \omega} \
\; \; \;\mbox{leading to} \; \; \;
\label{WFreltwomod}
\underline{\hat \alpha}_+ T \w  = (B - \mu \omega) \underline{\sigma}_+ -
\rho\,.
\ee
One way of interpreting (\ref{WFrelonemod}) and (\ref{WFreltwomod}) is
as Ward identities for $Q$ and $J$ correlators in a theory where
$\langle T_{aa} \rangle = P$ instead of $\calp$, thus shifting the
contact term subtraction required for $\langle TT \rangle_u$
and eliminating $mB$ from (\ref{eq:kappa}).

In \cite{HKMS}, the authors calculate $\underline \sigma$,
$\underline {\hat \alpha}$ and $\underline {\bar \kappa}$ using
the principles of magnetohydrodynamics (MHD). Their result for
$\underline \sigma$ is (\ref{MHDsigma}) but with a different
definition of the cyclotron pole
\be
\label{MHDcyclotronpole}
\omega_c = \frac{B \rho}{\epsilon+P} \; , \; \; \; \gamma = \frac{\sigma_Q B^2}{\epsilon+P} \ ,
\ee
where we have replaced $\calp$ with $P$.  For comparison with the MHD results,
we do not need an
explicit formula for $\sigma_Q$.
As it should be, the difference between these two conductivities is, to leading order
in $\omega$, our magnetization subtraction (\ref{WFreltwomod}):
\be
\underline \sigma_+ - \sigma_+ = \frac{m \omega}{B} + {\mathcal O}(\omega^2) \ .
\ee
That the higher order terms in $\omega$ do not match is not troubling because the
MHD result is only accurate to quadratic order in $\omega$.

Now given the MHD result $\underline \sigma_+$ for the conductivity,
along with the thermodynamic relation $\epsilon +P = s T + \mu \rho$ and the location of
the cyclotron pole (\ref{MHDcyclotronpole}),
one finds from (\ref{WFreltwomod}) that
\be
\underline{\hat \alpha}_+ =  \frac{s}{B} \frac{-\omega_c + i \gamma}{\omega + i \gamma - \omega_c}
- \frac{\omega }{T} \frac{i \sigma_Q \mu}{ \omega+ i \gamma - \omega_c} \ .
\ee
With a little more work, one also derives from (\ref{WFrelonemod}) that
\be
\underline{\bar \kappa}_+ T = \frac{- (sT)^2 +(i\mu \sigma_Q)  [ (\epsilon +P) (-B + \mu \omega) - B s T]}{(\epsilon + P) (\omega + i \gamma - \omega_c)}  \ .
\ee
These results match precisely the magnetization subtracted results from \cite{HKMS}.

We stress that we do not have any theoretical justificaton for
our subtractions away from the $\omega \to 0$ limit.
In the following, we compute $\sigma_+$ and not $\underline {\sigma}_+$,
and will not need to perform this subtraction. We note
also that the magnetization subtractions we include are all higher
order in the magnetic field. Our analytic formulae for $\sigma$
are only valid at leading order in $B$ and thus are insensitive to
these subtractions.

Having shown that the computation of $\hat \alpha$ and $\bar
\kappa$ reduces to that of finding the electrical conductivity
$\sigma$, we now compute $\sigma$ as a function of frequency for
the M2-brane theory. We will present both analytic and numerical
results.

\section{The low frequency limit}

\subsection{Hall conductivity}

Let us look first at stationary solutions where ${\mathcal{E}}$
and ${\mathcal{B}}$ are time independent. The equations of motion
immediately imply that
\be
{\mathcal{B}}_+ = -\frac{q}{h} {\mathcal{E}}_+ \,,
\ee
from which we have that
\be
\sigma_+ = - \frac{1}{g^2} \frac{q}{h} = \frac{\rho}{B} \quad \text{at} \quad \w=0 \,.
\label{hallconductivity}
\ee
In the last term we have expressed the result in terms of the field
theory charge density and background magnetic field using the
expressions above. Note that from (\ref{eq:xy}) we have that
$\sigma_+ = \sigma_{xy}$ in this case. Thus we recover the result of
\cite{Hartnoll:2007ai} for the hydrodynamic Hall conductivity,
as expected on general kinematic grounds.

\subsection{The hydrodynamic limit}
\label{sec:smallBrho}

In this section, we explore the equations (\ref{EBeqone}) and
(\ref{EBeqtwo}) in the limit where $w \to 0$ but $h^2 / w \equiv
H^2$ and $q^2/ w \equiv Q^2$ are held fixed.  We will find a pole
at the location predicted for the cyclotron frequency by
magnetohydrodynamics. In fact, we will recover precisely the
expressions derived from magnetohydrodynamics in \cite{HKMS}.

We begin by rewriting the two coupled first order equations
(\ref{EBeqone}) and (\ref{EBeqtwo}) as a second order equation for
${\mathcal E}_+$, suppressing the $+$ index for ease of notation:
\begin{eqnarray}
0 &=&  \cale''(z)  \left(w^2 - 4 h^2 z^2 f(z)\right)f(z)^2 +
\cale'(z) \left(f'(z) w^2 + 8 h^2 z f(z)^2\right) f(z) +  \nonumber \\
&&
      \cale(z) \left(w^4 - 8
      h^2 z^2 f(z) w^2 - 4 q^2 z^2 f(z)
        w^2 + 8 h q z f(z)^2 w +  \right. \nonumber \\
&&
\left.            4 h q z^2 f(z) f'(z) w + 16 h^4 z^4 f(z)^2 + 16 h^2 q^2 z^4
      f(z)^2\right) \,.
      \label{caleeq}
\end{eqnarray}
We next impose outgoing boundary conditions at the horizon $z=1$
by defining a new function $S(z)$,
\be
\cale(z) \equiv e^{i w \int_0^z dx/f(x)} S(z) \ ,
\ee
and imposing the constraint $S(1)=c_0$. As above, the time
dependence is of the form $e^{-i \omega t}$. We look for a series
solution to $S(z)$ of the form
\be
S(z) = S_0(z) + w S_1(z) + {\mathcal O}(w^2) \ .
\ee
The boundary condition at the horizon for $S(z)$ leads to the solutions
\be
S_0(z) =
c_0 \left(1  + \frac{4}{3} i H (H- i Q) (1-z^3) \right) \ ,
\ee
and
\be
S_1(z) =\frac{4}{3} i c_0 (H - i Q)^2 z (z - 1) (H z^2 (H + i Q)  + i (z + 1))\ .
\ee

To calculate the conductivity $\sigma_+$, we need to evaluate
$\cale$ and $\calb$ on the boundary $z=0$.  From (\ref{EBeqtwo}),
it is clear that on the boundary we have $w \calb(0) =
\cale'(0)$. In terms of the new function $S(z)$, $\cale'(0) =
S'(0) + i w S(0)$. Thus we find
\begin{eqnarray}\label{eq:MHD}
g^2 \sigma_+ &=& \frac{\calb(0)}{\cale(0)} = \frac{\cale'(0)}{ w \cale(0)}  \nonumber \\
&=&
\frac{S'(0)}{w S(0)} + i=
\frac{S_1'(0)}{S_0(0)} + i
=i \frac{4 i Q^2 - 4 H Q + 3}{4 i H^2 + 4 Q H + 3} \ .
\end{eqnarray}
There are several remarkable features of this expression.
First, in the low frequency limit $\omega \to 0$, the expression becomes
the standard formula for the Hall conductivity (\ref{hallconductivity}).

The second remarkable feature is the cyclotron frequency pole at
\be
w_* = -\frac{4}{3} h ( q + i h) \ .
\ee
In terms of the field theory variables, this pole is located at
\be
\omega_* = \omega_c - i \gamma =
\frac{\rho B}{ \epsilon + \calp} - i \frac{B^2}{g^2 (\epsilon+\calp)} \ ,
\ee
which is our expectation from magnetohydrodynamics
\cite{HKMS}, albeit we cannot in this limit tell the difference between
certain thermodynamic quantities. For example $sT \approx \epsilon
+ P$ and $\calp \approx P$. The fact that the pole has a negative imaginary part leads
to dissipation with the assumed time dependence $e^{-i \omega t}$.
One way of understanding the normalization of the decay is the
fact that $q+ih$ vanishes for self-dual configurations in the
bulk, with $q = -ih$.  Such a self-dual field strength does not back
react on the geometry and thus must lead to the
frequency independent conductivity found in \cite{HKSS}.

The third interesting feature of this expression is the way that
it realises the expected symmetry under S duality: $q
\to h$ and $h \to -q$. The expression for $\sigma$ has a zero
in precisely the right place to become a pole for the S dual
expression.

In field theory variables, the complexified conductivity can be written
\be
\sigma_+ = \frac{i}{g^2} \frac{\omega + i \omega_c^2/\gamma + \omega_c}
{\omega + i \gamma - \omega_c} \ .
\ee
In the limit we are working, we cannot tell the difference between
$1/g^2$ and $\sigma_Q$.  Nor can we tell the difference between
$\epsilon+\calp$ in the location of the cyclotron pole
(\ref{cyclotronpole}) and $sT$.  However, in the next section, in
which we investigate small $h$ and arbitrary $q$, we will find
additional constraints on the way $q$ must appear in this
expression which is consistent with (\ref{MHDsigma}).

\subsection{Small magnetic field}
\label{sec:smallB}

In this section, we consider the small frequency and magnetic field
limit in which $H = h/w$ is held constant.
The results will allow us to expand the validity of the hydrodynamic
limit to finite $q$.
In this limit, (\ref{caleeq}) becomes
\begin{eqnarray}
 \label{caleeqsmallh}
0 &=&  \cale''(z)  \left(1 - 4 H^2 z^2 f(z)\right)f(z) +
\cale'(z) \left(f'(z) + 8 H^2 z f(z)^2\right)  +  \\
&&
      \cale(z) \left(- 4 q^2 z^2
         + 8 H q z f(z)  +
            4 H q z^2  f'(z)   + 16 H^2 q^2 z^4
      f(z)\right) + {\mathcal O}(w^2) \,. \nonumber
\end{eqnarray}
where now $f(z) = 1 -(1+q^2) z^3 + q^2 z^4$.

We can solve this differential equation exactly to find
\be
\cale = \left( \frac{f'}{z^2} - 4 H q f \right) \left( c_1 + c_2 \int_0^z
\frac{dx}{f(x)} \frac{1-4 H^2 x^2 f(x)}{\left( \frac{f'(x)}{x^2} - 4 H q f(x) \right)^2}
\right) \ .
\ee

Close to the horizon, we impose outgoing boundary conditions
\be
\cale \sim (1-z)^{iw / (q^2-3)} = 1 + \frac{iw}{q^2-3} \ln (1-z) + \ldots \
\ee
and these boundary conditions impose a relation between $c_1$ and $c_2$.
We find
\be
\frac{c_2}{c_1} = i (q^2-3)^2 w + {\mathcal O}(w^2)\ .
\ee

The conductivity can now be computed using (\ref{eq:conductivity}),
\be
g^2 \sigma_+ = \frac{\cale'(0)}{w \cale(0)} =
- \frac{4 q^2}{w(3+4Hq+3q^2)} + i\frac{(3-q^2)^2}{(3+4Hq+3q^2)^2} +
{\mathcal O}(w)\ .
\ee
We can rewrite this conductivity in field theory variables (in the limit where $h$ is small)
yielding
\be
\label{smallhsigma}
\sigma_+ = - \frac{\rho^2}{(\epsilon + \calp)(\omega-\omega_c) } +
i \frac{\sigma_Q \omega^2}{(\omega-\omega_c)^2}
\ee
where we have defined
\be
\label{sigmaQ}
\sigma_Q \equiv \left(\frac{3-q^2}{3 (1+q^2)} \right)^2 \frac{1}{g^2} =
\left( \frac{sT}{\epsilon+\calp}\right)^2 \frac{1}{g^2}\ .
\ee
Expanding (\ref{MHDsigma}) in the limit of small $h$ and $w$ with
$h/w$ held fixed yields precisely (\ref{smallhsigma}). The
computation of this section has allowed us to determine the $q$
dependence of $\sigma_Q$. A similar result for $\sigma_Q$ in the
$AdS_5$ case was presented in \cite{Son:2006em}.

We should emphasize that the higher order dependence on $h$ of
$\omega_*$ and $\sigma_Q$ in (\ref{cyclotronpole}) and
(\ref{sigmaQ}) is nothing more than an inspired guess. Large $h$
takes us out of the small $\omega$ frequency regime where we have
analytic control.

\subsection{A semicircle law}

It is interesting to consider the case of vanishing charge, $q=0$.
The hydrodynamic conductivity (\ref{eq:MHD}) becomes
\be
g^2 \sigma_+ = \frac{1}{4 H^2/3 - i} \,.
\ee
As we vary $H$, this conductivity obeys
\be
\left|g^2 \sigma_+ - \frac{i}{2} \right| = \frac{1}{2} \,.
\ee
Thus, the conductivity
traces out a semicircle in the complex conductivity plane from the
insulator $\sigma_+ = 0$ to $g^2 \sigma_+ = i$.
Given that $H^2$ is always positive, we only obtain the half of
a full circle with positive real part.

Because $h$ appears quadratically in the conductivity, it follows
from (\ref{eq:xy}) that $\sigma_{xy} = 0$. Thus the conductivity
in this case is purely diagonal: $\sigma_+ = i
\sigma_{xx}$. This semicircle therefore does not appear to be
related to the semicircle laws that are observed in the
transitions between quantum Hall plateaux \cite{Hilke} and conjectured
to be related to subsets of $SL(2,{\mathbb Z})$ invariance
\cite{DykhneRuzin, LutkenRoss, BurgessDolan}.
Rather, the origin of our semicircle law can be traced to a
general feature of any resonance.

Near a pole $z = z_0$, the expression for the conductivity has the general form
\be
\sigma(z)  \sim \frac{a}{z-z_0} \ .
\label{approxcond}
\ee
If we plot parametrically $\{\mbox{Re}(\sigma(x)), \mbox{Im}(\sigma(x))\}$
as a function of $x$ for $-\infty < x < \infty$, then the curve traces out a circle
that passes through the origin $z=0$ and is centered around $z = -a / (z_0-\bar z_0)$.
The smaller $\mbox{Im}(z_0)$, the better $\sigma(z)$ is approximated by
(\ref{approxcond}) near $x= \mbox{Re}(z_0)$, and the more circular the
parametric plot.

\begin{figure}
\centerline{a) \epsfig{figure=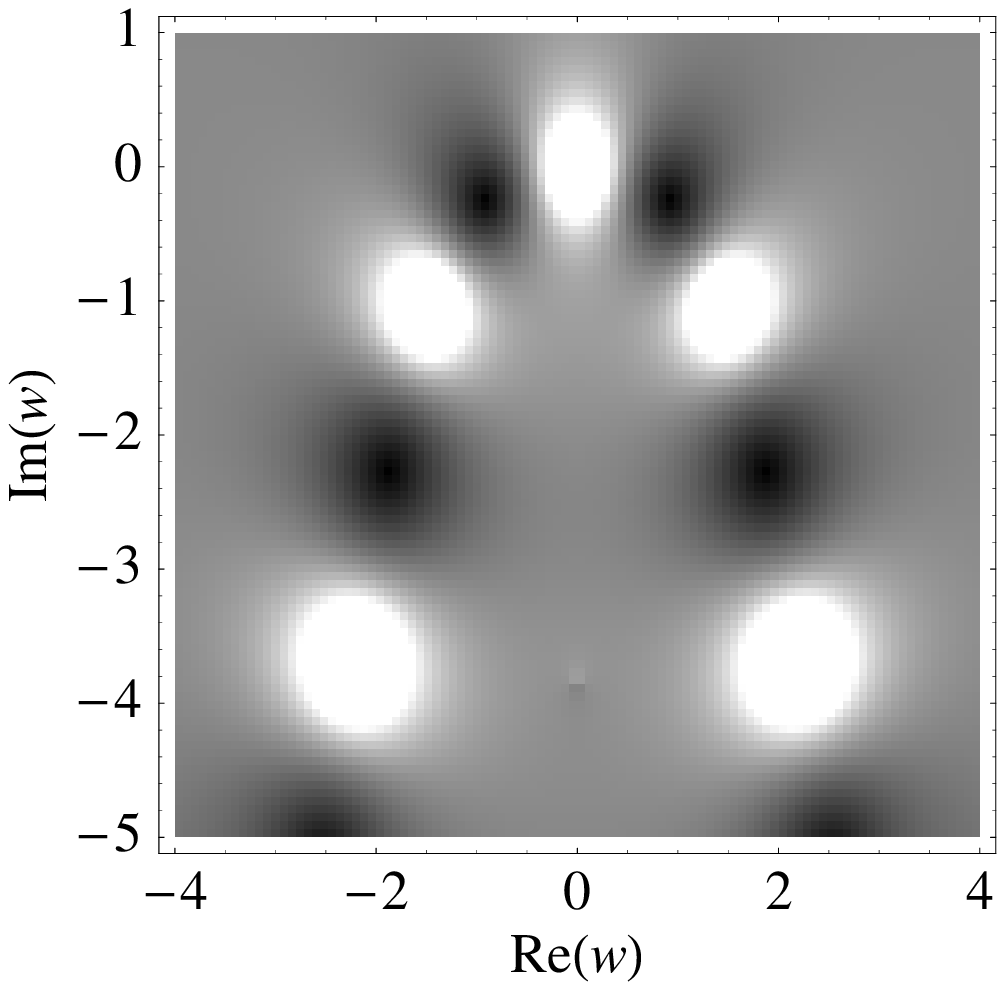, width=2in}  b) \epsfig{figure=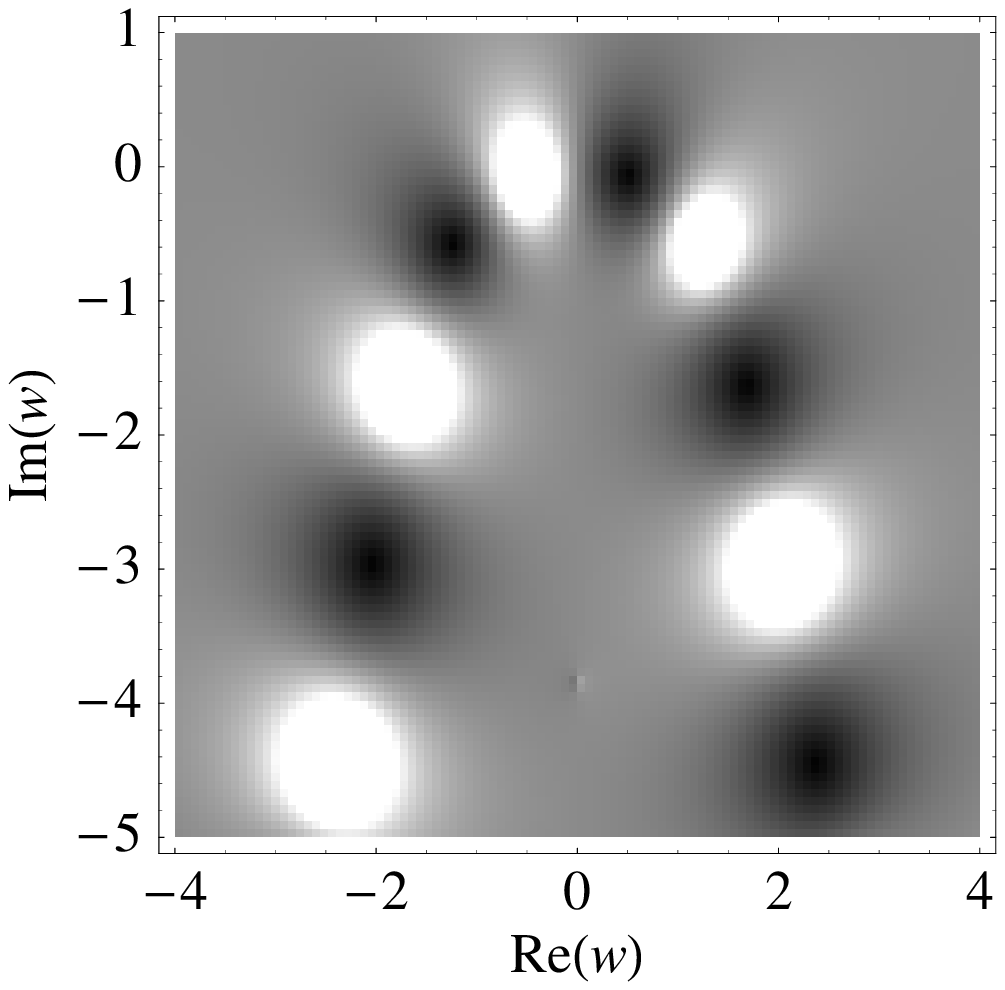, width=2in} c) \epsfig{figure=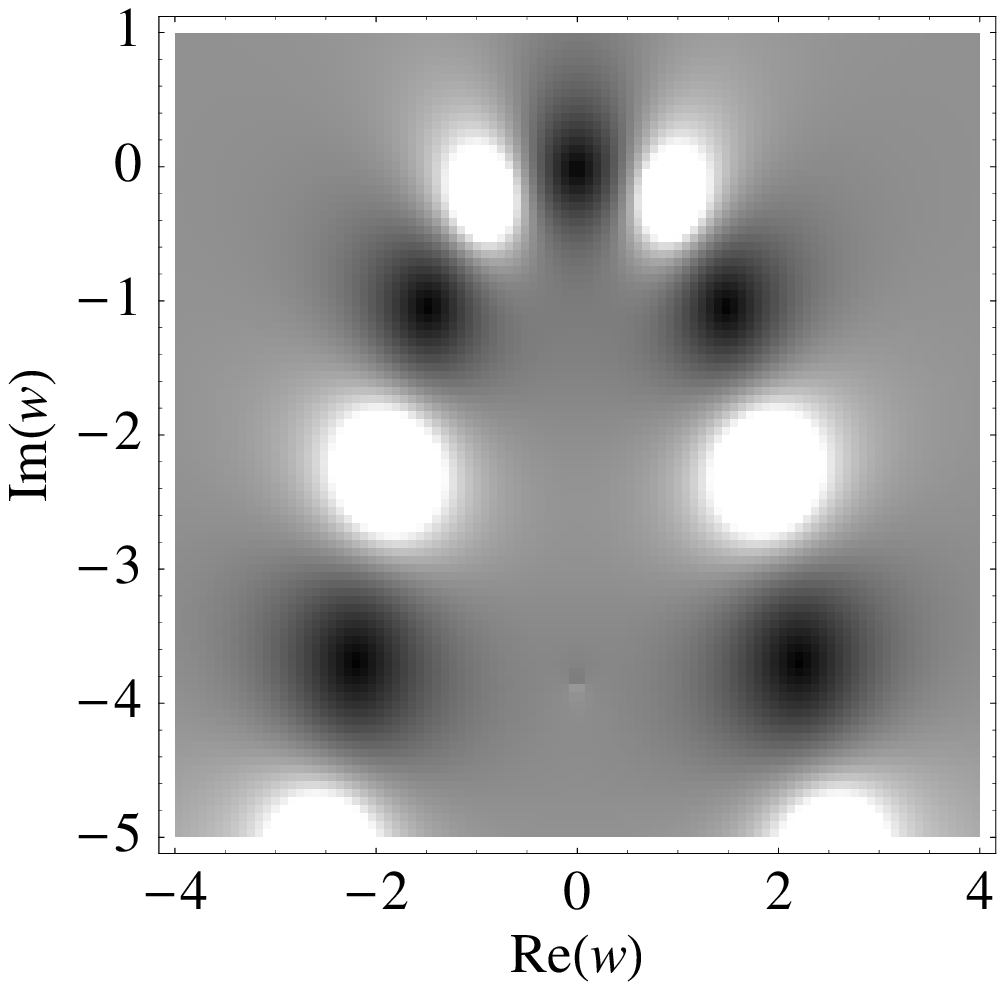, width=2in}}
\caption{
\label{densityplots}
A density plot of $|\sigma_+|$ as a function of complex $w$.  White
areas are large in magnitude and correspond to poles while dark
areas are zeroes of $\sigma_+$: a) $h=0$ and $q=1$, b)
$h=q=1/\sqrt{2}$, c) $h=1$ and $q=0$.}
\end{figure}

\section{The cyclotron resonance at general frequency}
\label{sec:numerics}

\begin{figure}
\centerline{a) \epsfig{figure=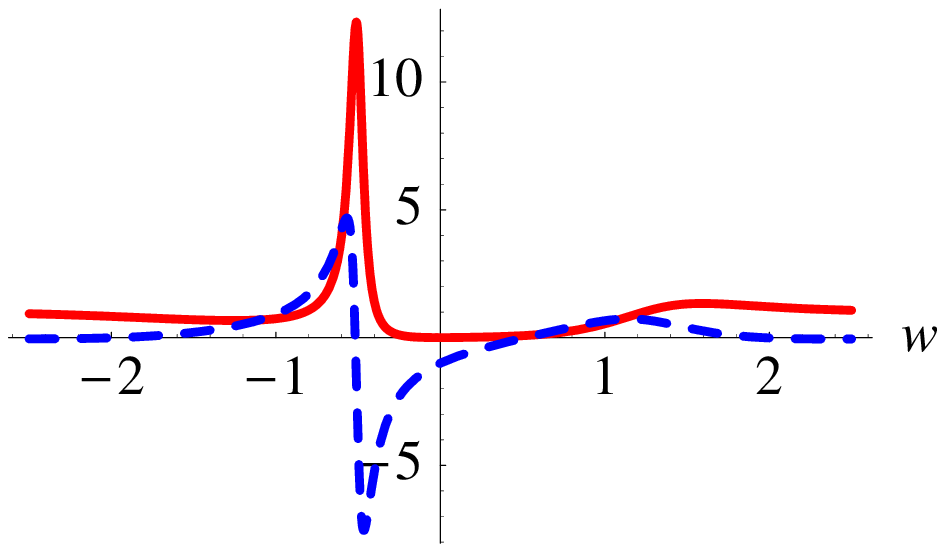, width=2.5in}  b) \epsfig{figure=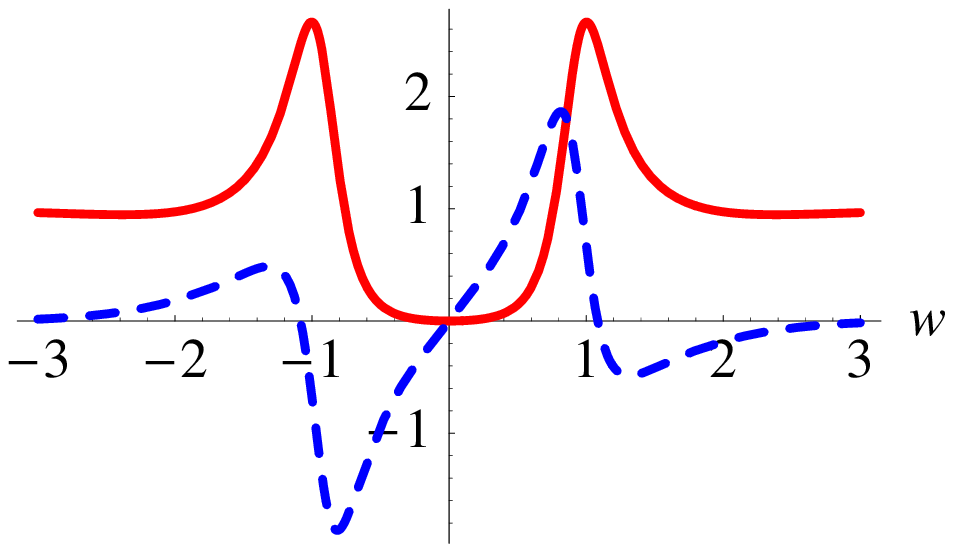, width=2.5in}}
\caption{
\label{realsliceplots}
The dashed blue line is the $\mbox{Im}(\sigma_+)$ while the solid
red line is the $\mbox{Re}(\sigma_+)$ as a function of $w$: a)
$h=q=1/\sqrt{2}$, b) $h=1$ and $q=0$. }
\end{figure}

\begin{figure}
\centerline{a) \epsfig{figure=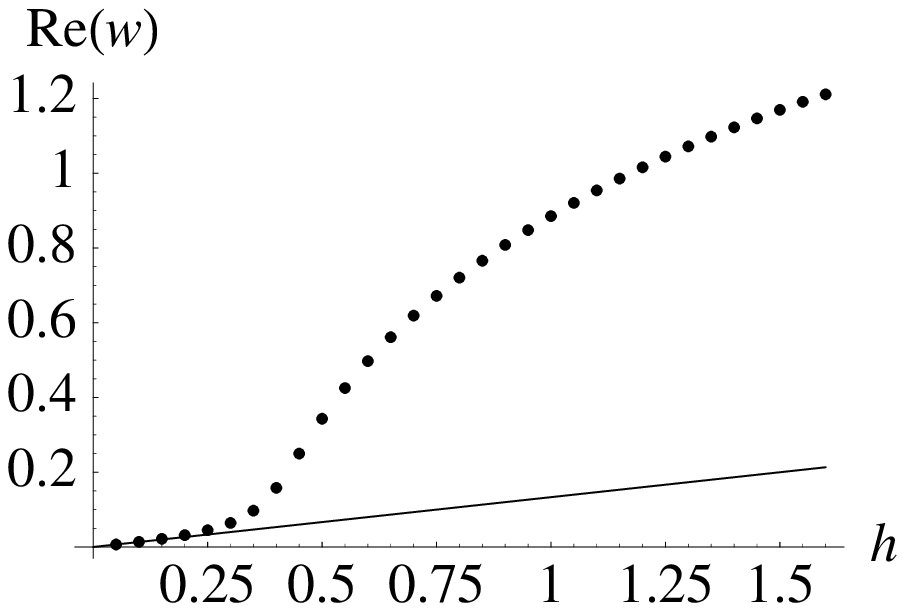, width=3in} b) \epsfig{figure=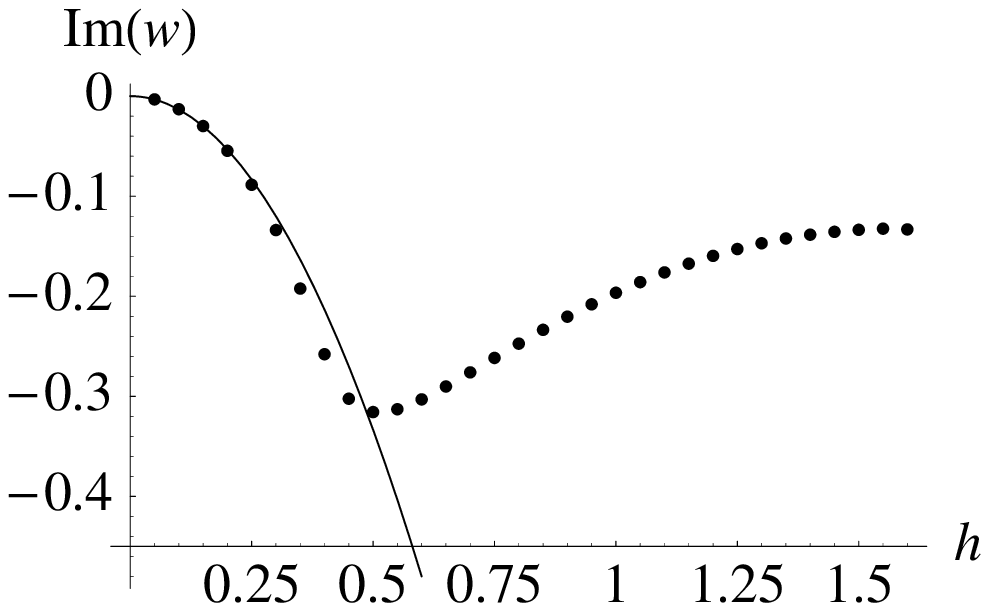, width=3in}}
\centerline{c) \epsfig{figure=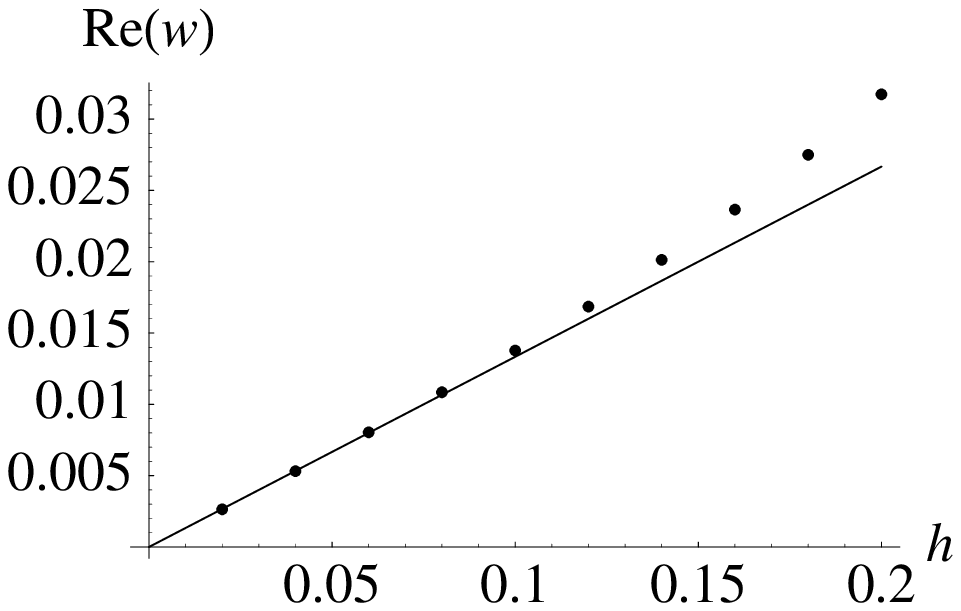, width=3in}  d) \epsfig{figure=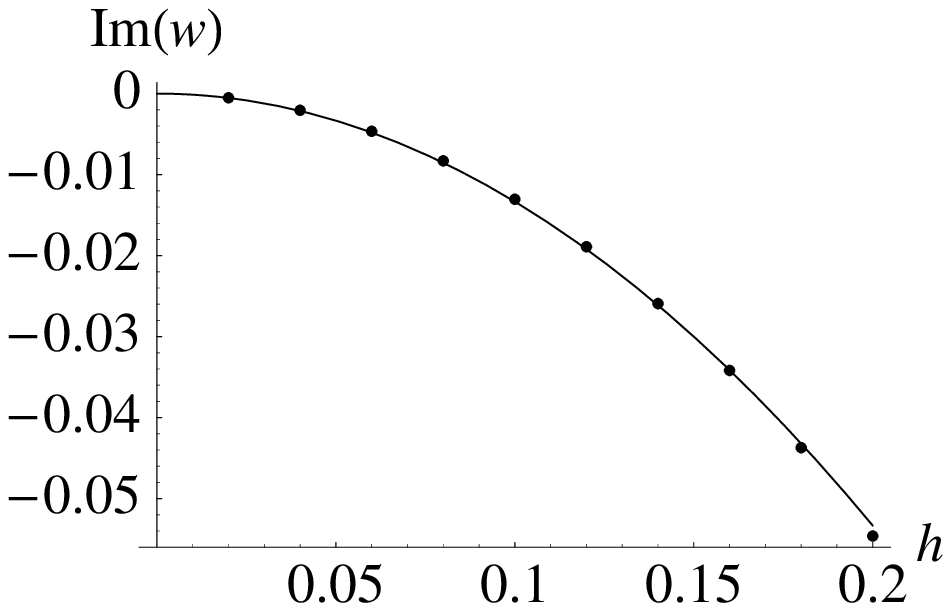, width=3in}}
\caption{
\label{hydroplots}
The location of the pole closest to the origin as a function of
$h$ for $q=-0.1$.  The data points are numerically determined
locations of the pole.  The curves show the limiting hydrodynamic
behavior.  Plots (c) and (d) are closeups of the hydrodynamic
region in plots (a) and (b).}
\end{figure}

\begin{figure}
\centerline{a) \epsfig{figure=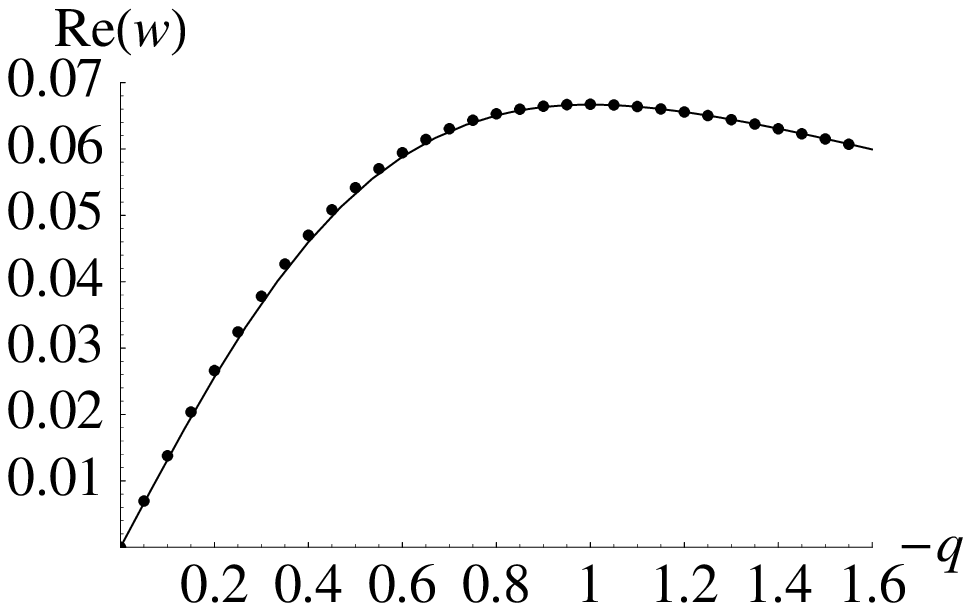, width=3in} b) \epsfig{figure=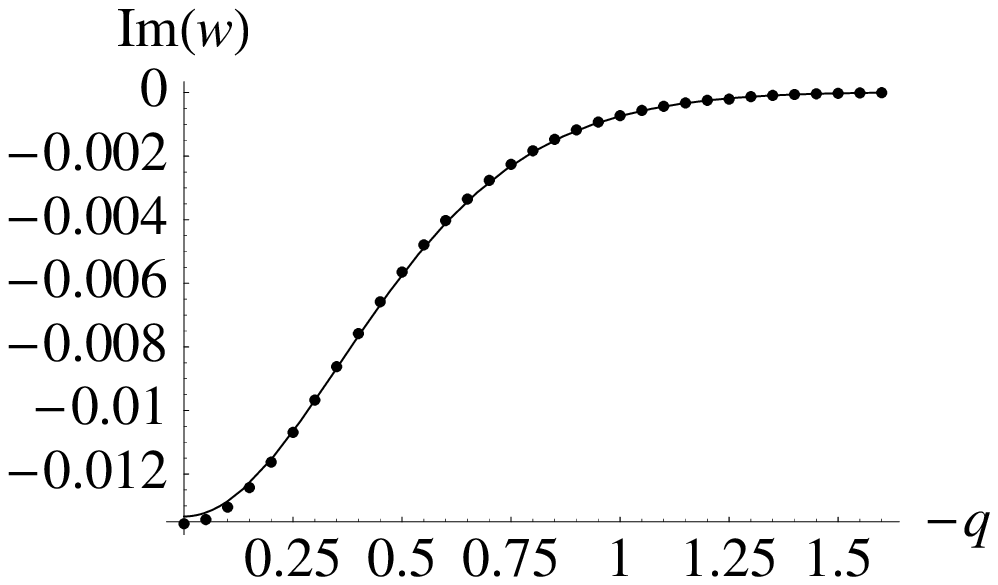, width=3in}}
\caption{
\label{hydroqplots}
The location of the pole closest to the origin as a function of
$-q$ for $h=0.1$.  The data points are numerically determined
locations of the pole.  The curves show the limiting hydrodynamic
behavior.}
\end{figure}

We have not succeeded in finding a general analytic solution for
(\ref{EBeqone}) and (\ref{EBeqtwo}), but we were able to solve the
first order system numerically. In this section we present
numerical results for the complexified conductivity as a function
of frequency, not necessarily small. Furthermore, we will trace
the motion of the cyclotron resonance pole in the complex
frequency plane as a function of $B$ and $\rho$.
The results are shown in figures \ref{densityplots},
\ref{realsliceplots}, \ref{hydroplots}, and \ref{hydroqplots}.

A small difficulty in the numerical integration is
applying outgoing boundary conditions at the horizon.  Canned
differential equation solvers such as Mathematica's NDSolve
typically require initial conditions to be specified as the value
of a function and its first few derivatives at a point. However,
the horizon is a singular point in the differential equation. To
enforce the outgoing boundary conditions at the horizon, we solved
analytically for the first few terms of a power series solution in
$z-1$ for $S(z)$ near the horizon.  Then we started the numerical
differential equation solver a small distance $\epsilon$ away from
the horizon where the differential equation is regular.

In figure \ref{densityplots}, we present a three dimensional plot
of $|\sigma_+|$ as a function of the complexified frequency
$\omega$ for different values of $h$ and $q$.  As we shift the
values of $h$ and $q$, holding $h^2+q^2$ constant, the locations
of the poles and zeroes of $\sigma$ shift around the arch-like
configuration. The fact that figure \ref{densityplots}a is a
photographic negative of figure \ref{densityplots}c is a
consequence of S duality. As we alter the ratio of $h$ and $q$ the
poles and zeroes rotate, until they have precisely exchanged
location at the dual value. The slice along the real axis of
figures \ref{densityplots}b and \ref{densityplots}c is shown as
figures \ref{realsliceplots}a and \ref{realsliceplots}b
respectively.

In figure \ref{hydroplots}, we investigate the location of the
cyclotron pole $\omega_*$ in the limit of small $\rho$
and arbitrary $B$.  When both $h$ and $q$ are small, the location
of the pole is well approximated by the formulae (\ref{cyclotronpole})
which were valid precisely when $\omega_*$ was small.
However, as $h$ increases, $\omega_*$ increases as well and the
increase eventually takes us out of the regime where (\ref{cyclotronpole}) is valid.

In figure \ref{hydroqplots}, we look at $\omega_*$ in the limit of
small $B$ and arbitrary $\rho$.  In this limit, $\omega_*$ is
always small and the corresponding formulae (\ref{cyclotronpole})
are expected to be valid always.  Indeed, the matching between the
numeric and analytic result is remarkably good for all $q$ at
$h=0.1$. This agreement confirms that the combination of limits we
considered in the previous section have correctly captured the
dependence of the cyclotron pole on arbitrary $\rho$ with small
$B$.

\section{Discussion}

In this paper we have used the AdS/CFT correspondence to study
thermoelectric transport in a strongly coupled conformal field
theory at finite temperature, electric charge density and
background magnetic field. By solving the equations for
perturbations about the dual dyonic black hole background, we
obtained a combination of analytic and numerical results for the
electrical conductivity. We have then shown that this conductivity
determines the other thermoelectric transport coefficients of the
CFT.

There are two important qualitative features of our results. The
first is the existence of relativistic, damped cyclotron
resonances due to the background magnetic field. These resonances
lead to important features in the conductivity as a function of
frequency, as in figure \ref{realsliceplots}. We have explicitly
exhibited this resonance as a pole in the complex frequency plane;
analytically for small magnetic fields and numerically for general
values of the magnetic field. When this pole comes close to the
real frequency axis, it can result in semicircle laws for the
complexified conductivities as a function of the magnetic field or
the frequency.

The second important feature is that electromagnetic duality of
the bulk theory acts nontrivially on the transport coefficients of
the CFT. In field theory this duality exchanges the values of the
background magnetic field and the charge density, and can be
thought of as a particle-vortex duality
\cite{HKSS, Witten:2003ya}. Under this exchange, we have shown
that the complexified conductivity transforms as $2
\pi \sigma \to -1/2 \pi \sigma$. Thus the duality constrains the
dependence of the conductivity on the magnetic field and charge
density. Looking at the conductivity in the complex frequency
plane, in figure \ref{densityplots}, we see an interesting pattern
of poles and zeros that are exchanged under duality. Although
exact self-duality is a special feature of CFTs with an Anti-de
Sitter dual described by Einstein-Maxwell theory, in the
hydrodynamic limit our expressions precisely match generic
magnetohydrodynamic (MHD) expectations \cite{HKMS}. Thus we obtain
a dual understanding of why relativistic MHD results for the
transport coefficients exhibit an interesting and perhaps
unexpected duality.

One motivation for our work was to make a connection with the
physics of quantum critical phenomena in 2+1 dimensional condensed
matter systems. Finite temperature conformal field theories are
the appropriate description of such systems when the temperature
is the most important scale near the critical point. One example
of such critical phenomena is the vicinity of a
superfluid-insulator transition in cuprate superconductors. Recent
measurements of the Nernst effect in this regime \cite{Ong}
require better theoretical models. The Nernst coefficient measures
the transverse voltage arising due to both a thermal gradient and
a background magnetic field, and can be computed from $\hat
\alpha$ and $\sigma$. An MHD calculation of the Nernst coefficient was
presented in detail in \cite{HKMS}, and some agreement with
measurements achieved. Furthermore, the MHD analysis leads to a
prediction of a cyclotron resonance which could be observed in the
future. In this paper we have derived using the AdS/CFT
correspondence all of the formulae that are used in such MHD
computations.

Various extensions of our work are possible. For instance, it
would be interesting to generalize our MHD formulae to include
finite spatial momentum. It would also be of interest to apply our
formalism to other asymptotically $AdS_4$ backgrounds, such as
those arising in flux compactifications. Indeed,
\cite{Karch:2007pd} already studies how Ohm's law emerges for
systems of D-brane probes where the separation of scales between
the bulk and D-brane degrees of freedom leads to a finite
conductivity at $\rho \neq 0$ and $B=0$, in contrast to the
results here where translation invariance guarantees the
conductivity diverges in this limit.  To get a finite conductivity
at $B=0$, in another possible extension of this work, we would
need to find a holographic way of adding disorder to our system.
Finally, we would like to know if there are physical systems other
than those discussed in \cite{HKMS} where relativistic MHD is an
appropriate description. So far, we do not see a direct connection
between the standard AdS/CFT setups and the (fractional) quantum
Hall effect, but any such connection would be fascinating.

\section*{Acknowledgments}

We would especially like to thank Andreas Karch for many helpful
comments and collaboration in the early stages of this project. It
is also a pleasure to acknowledge conversations with David Berman,
Cliff Burgess, Pavel Kovtun, Markus M\"uller, Andy O'Bannon,
Subir Sachdev, Dam
Son, and Larry Yaffe. Both authors acknowledge the hospitality of
the Perimeter Institute where part of this work was completed.
This research was supported in part by the National Science
Foundation under Grant No. PHY05-51164 and in part by the U.S.
Department of Energy under Grant No.~DE-FG02-96ER40956.

\appendix

\end{document}